\documentclass[useAMS]{mn2e}
\pdfoutput=1
\usepackage{epsfig}
\usepackage{amsmath}

\def\be{\begin{equation}}
\def\ee{\end{equation}}
\def\ba{\begin{eqnarray}}
\def\ea{\end{eqnarray}}
\def\go{\mathrel{\raise.3ex\hbox{$>$}\mkern-14mu
             \lower0.6ex\hbox{$\sim$}}}
\def\lo{\mathrel{\raise.3ex\hbox{$<$}\mkern-14mu
             \lower0.6ex\hbox{$\sim$}}}
\def\bxi{{\mbox{\boldmath $\xi$}}}
\def\br{{\bf r}}
\def\bD{{\bf D}}

\def\min{{\rm min}}
\def\max{{\rm max}}
\def\mratio{\left(\frac{M'}{M}\right)}
\def\mtratio{\left(\frac{M_t}{M}\right)}
\def\bomega{{\bar\omega_\alpha}}
\def\bQ{{\bar Q_\alpha}}

\begin{document}

\title[Resonant Oscillation Modes in White Dwarf Binaries]
{Tidal Excitations of Oscillation Modes in Compact White Dwarf Binaries: I. Linear Theory}
\author[J. Fuller and D. Lai]
{Jim Fuller\thanks{Email:
derg@astro.cornell.edu; dong@astro.cornell.edu}
and Dong Lai\\
Center for Space Research, Department of Astronomy, Cornell University, Ithaca, NY 14853, USA}

\label{firstpage}
\maketitle

\begin{abstract}
We study the tidal excitation of gravity modes (g-modes) in compact
white dwarf binary systems with periods ranging from minutes to hours.
As the orbit of the system decays via gravitational radiation, the
orbital frequency increases and sweeps through a series of resonances
with the g-modes of the white dwarf. At each resonance, the tidal
force excites the g-mode to a relatively large amplitude, transferring
the orbital energy to the stellar oscillation. We calculate the
eigenfrequencies of g-modes and their coupling coefficients with the
tidal field for realistic non-rotating white dwarf models. Using these
mode properties, we numerically compute the excited mode
amplitude in the linear approximation as the orbit passes though the
resonance, including the backreaction of the mode on the orbit.  We
also derive analytical estimates for the mode amplitude and the duration
of the resonance, which accurately reproduce our numerical results for most 
binary parameters. We find that the g-modes can be excited to 
a dimensionless (mass-weighted) amplitude up to 0.1, with the mode energy 
approaching $10^{-3}$ of the gravitational binding energy of the star. Therefore the low-frequency ($\lo 10^{-2}$~Hz) gravitational waveforms
produced by the binaries, detectable by LISA, are strongly affected by the tidal resonances.
Our results also suggest that thousands of years prior to the binary merger, the 
white dwarf may be heated up significantly by tidal interactions.
However, more study is needed since the physical amplitudes of the excited
oscillation modes become highly nonlinear in the outer layer of the star,
which can reduce the mode amplitude attained by tidal excitation.
\end{abstract}


\section{Introduction}

It is well known that non-radial gravity modes (g-modes) 
are responsible for the luminosity variations observed in some 
isolated white dwarfs (called ZZ Ceti stars) in the instability strip.  
These g-modes are thought to be excited by a convective driving mechanism
operating in the shallow surface convection zone of the star 
(see Brickhill 1983; Goldreich \& Wu 1999; Wu \& Goldreich 1999).

In this paper we study the tidal excitation of g-modes in compact
binary systems containing a white dwarf (WD) and another compact
object (white dwarf, neutron star or black hole).  The Galaxy is
populated with $\sim 10^8$ WD-WD binaries and several $10^6$ of double
WD-NS binaries 
[Nelemans et al.~(2001); see also Nelemans (2009) and references therein].  
A sizeable fraction of these binaries are compact enough so that the
binary orbit will decay within a Hubble time to initiate mass transfer
or a binary merger.  Depending on the details of the mass transfer
process (including the response of the WD to mass transfer), these
ultra-compact binaries (with orbital period less than an hour) may
survive mass transfer for a long time or merge shortly after mass
transfer begins. A number of ultra-compact interacting WD-WD binary
systems have already been observed [including RX J0806.3+1527 (period
5.4 min) and V407 Vul (period 9.5 min); see Strohmayer 2005 and
Ramsay et al.~2005].  Recent surveys (e.g., SDSS) have also begun
to uncover non-interacting compact WD binaries (e.g., Badenes et al.~2009; 
Mullally et al.~2009;
Kilic et al.~2009; Marsh et al.~2010; Kulkarni \& van Kerkwijk 2010;
Steinfadt et al.~2010).
Depending on the total mass, the systems may evolve into Type Ia supernovae
(for high mass), or become AM CVn binaries or R CrB stars (for low mass).
Many of these WD binaries are detectable in gravitational waves by 
the {\it Laser Interferometer Space Antenna (LISA)} (Nelemans 2009).

In this paper we consider resonant tidal interaction in WD binaries
that are not undergoing mass transfer. This means that the binary
separation $D$ is greater than $D_\min$, the orbital radius at which
dynamical merger or mass transfer occurs, i.e.,
\be
\label{tidallimit}
D\go D_{\min}\simeq 2.5\,\mtratio^{1/3}R,
\ee
where $M$ is the WD mass, $M_t=M+M'$ is the total mass and $R$ is the 
WD radius. This corresponds to orbital periods of
\be
P\go P_{\min}= 68.4 \ \bigg(\frac{R}{10^4 \textrm{km}}\bigg)^{3/2} \bigg(\frac{M}{M_{\odot}}\bigg)^{-1/2} \textrm{s}.
\ee
Since WD g-mode periods are of order one minute or longer, they can be excited
by the binary companion prior to mass transfer. In particular, as the
binary orbit decays due to gravitational radiation, the orbital frequency
sweeps through a series of g-mode frequencies, transferring orbital energy to
the modes. Although the overlap integral of the g-mode eigenfunctions with the 
tidal potential is generally quite small, a binary system that spends a long time 
at resonance can still excite g-modes to large amplitudes.

Previous studies of tidal interaction in WD binaries have focused on
quasi-static tides (e.g. Iben, Tutukov \& Fedorova 1998; Willems,
Deloye \& Kalogera 2009), which essentially correpond to non-resonant
f-modes of the star.  Such static tides become important only as the binary
approaches the tidal limit [equation (\ref{tidallimit})].  Racine,
Phinney \& Arras (2007) recently studied non-dissipative tidal
synchronization due to Rossby waves in accreting ultra-compact WD
binaries.  Rathore, Blandford \& Broderick (2005) studied resonant
mode excitations of WD modes in eccentric binaries. They focused on
f-modes, for which the resonance occurs when harmonics of the orbital
frequency matches the mode frequency. As mentioned above, for circular
orbits, such resonance with the f-mode does not occur prior to mass
transfer or tidal disruption. Their published analysis also did not
include back reaction of the excited mode on the binary orbit.

The problem of resonant mode excitations in compact binaries has been
studied before in the context of coalescing neutron star binaries:
Reisenegger \& Goldreich (1994), Lai (1994) and Shibata (1994)
focused on the excitations of g-modes of non-rotating neutron stars;
Ho \& Lai (1999) and Lai \& Wu (2006) studied the effects of NS
rotation -- including r-modes and other inertial modes; 
Flanagan \& Racine (2006) examined gravitomagnetic excitation of r-modes. 
In the case of neutron star binaries, the orbital decay
rate (for orbital frequencies larger than 5 Hz) 
is large and the mode amplitude is rather small, so the back
reaction of the excited mode on the orbit can be safely neglected
(see section 5 of the present paper).  By contrast, in the case of
WD binaries, the orbital decay is much slower and the excited mode
can reach a much larger amplitude. It thus becomes essential to take
the back reaction into account.

In this paper, we consider WD binaries in circular orbits, consistent with the observed population of compact WD binaries (e.g. Kulkarni \& van Kerkwijk 2010). Such circular orbits are a direct consequence of the circularization by gravitational radiation and/or the common envelope phase leading to their formation. A key assumption of this paper is that we assume the WD is not
synchronized with the binary orbit. While it is true that the tidal
circularization time scale is much longer than the synchronization
time, the observed circular orbit of the WD binaries does not imply synchronization. While there have been numerous studies of tidal
dissipation in normal stars and giant planets 
(e.g., Zahn 1970,1989; Goldreich \& Nicholson 1977; Goodman \& Oh 1997;
Goodman \& Dickson 1998;
Ogilvie \& Lin 2004,2007; Wu 2005; Goodman \& Lackner 2009),
there has been no satisfactory study on
tidal dissipation in WDs. Even for normal stars, the problem is not
solved (especially for solar-type stars; see Goodman \& Dickson 1999;
see Zahn 2008 for review). In fact it is likely 
that the excitations of g-modes and other low-frequency modes
play a role in the synchronization process.  The orbital decay time scale
near g-mode resonances is relatively short (of order $10^4$ years for orbital
periods of interest, i.e., minutes), 
so it is not clear that tidal synchronization
can compete with the orbital decay rate.
Given this uncertainty, we will consider non-rotating WDs 
(or slowly-rotating WDs, so that the g-mode properties are not 
significantly modified by rotation) as a first
step, and leaving the study of the rotational effects to a future paper.

This remainder of the paper is organized as follows. In section 2 we
present the equations governing the evolution of the orbit and the
g-modes. Section 3 examines the properties of WD g-modes and their
coupling with the tidal gravitational field of the companion. In
section 4, we numerically study the evolution of the g-modes through
resonances, and in section 5 we present analytical estimates of the
resonant g-mode excitation. We study the effect of mode damping on
the tidal excitation in section 6 and discuss the uncertainties and
implications of our results in section 7.

\section{Combined Evolution Equations for Oscillation Modes and Binary Orbit}

We consider a WD of mass $M$ and radius $R$ in orbit with 
a companion of mass $M'$ (another WD, or NS or BH). The WD is 
non-spinning. The gravitational potential produced by $M'$ can be written as
\begin{align}
U(\br,t) &= -\frac{GM'}{|\br-\bD(t)|}\nonumber\\
 &= -GM'\sum_{lm}\frac{W_{lm}r^l}{D^{l+1}}\,\, e^{-im\Phi(t)}
Y_{lm}(\theta,\phi),
\end{align}
where $\br=(r,\theta,\phi)$ is the position vector (in spherical
coordinates) of a fluid element in star $M$,
$\bD(t)=(D(t),\pi/2,\Phi(t))$ is the position vector of $M'$ relative to $M$
($D$ is the binary separation, $\Phi$ is the orbital phase or the true
anomaly) and the coefficient $W_{lm}$ is given by
\begin{align}
W_{lm} &= (-)^{(l+m)/2}\bigg[\frac{4\pi}{2l+1}(l+m)!(l-m)!\bigg]^{1/2}\nonumber\\
&\quad \times \bigg[2^l \bigg(\frac{l+m}{2}\bigg)!\bigg(\frac{l-m}{2}\bigg)!\bigg]^{-1},
\end{align}
(Here the symbol $(-)^p$ is zero if $p$ is not an integer.)
The dominant $l=2$ tidal potential 
has $W_{2\pm 2}=(3\pi/10)^{1/2}$, $W_{20}=(\pi/5)^{1/2}$,
$W_{2\pm 1}=0$, and so only the $m=\pm 2$ modes can be resonantly excited.

The linear perturbation of the tidal potential on $M$ is specified
by the Lagrangian displacement $\bxi(\br,t)$, which satisfies the
equation of motion
\be
\frac{\partial^2 \bxi}{\partial t^2}+{\cal L}\cdot\bxi=-\nabla U,
\ee
where $\cal L$ is an operator that specifies the internal restoring forces of
the star. The normal oscillation modes of the star satisfy
${\cal L}\cdot\bxi_\alpha=\omega_\alpha^2\bxi_\alpha$, 
where $\alpha=\{n,l,m\}$ is the usual mode index and 
$\omega_\alpha$ is the mode frequency.
We write $\bxi(\br,t)$ as the sum of the normal modes:
\be
\bxi(\br,t)=\sum_\alpha a_\alpha(t)\bxi_\alpha(\br).
\ee
The (complex) mode amplitude $a_\alpha(t)$ satisfies the equation
\be
\ddot a_\alpha+\omega_\alpha^2 a_\alpha=\frac{GM'W_{lm}Q_\alpha}{D^{l+1}}
\,e^{-im\Phi(t)},
\label{eq:addot}
\ee
where $Q_\alpha$ is the tidal coupling coefficient (also used by Press \& Teukolski 1977), defined by
\begin{align}
Q_\alpha &= \langle \bxi_\alpha | \nabla (r^l Y_{lm}) \rangle \nonumber \\
 &= \int\!d^3x\,\,\rho\bxi_\alpha^\ast\cdot \nabla (r^lY_{lm}) \nonumber \\
 &= \int\!d^3x\,\,\delta\rho_\alpha^\ast\, r^l Y_{lm}.
\label{eqq}
\end{align}
Here $\delta\rho_\alpha=-\nabla\cdot(\rho\bxi_\alpha)$ is the Eulerian density
perturbation. In deriving (\ref{eq:addot}) we have used the normalization
\be
\langle \bxi_\alpha | \bxi_\alpha \rangle = \int\!d^3x\,\rho\,\bxi_\alpha^\ast\cdot\bxi_\alpha=1.
\ee
Resonant excitation of a mode $\alpha$ occurs when $\omega_\alpha = m \Omega$, where $\Omega$ is the orbital frequency.

In the {\it absence} of tidal interaction/resonance, the WD binary orbit
decays due to gravitational radiation, with time scale given by (Peters 1964)
\begin{align}
t_D &= \frac{D}{|\dot{D}|} = \frac{5c^5}{64G^3}\frac{D^4}{MM'M_t}\nonumber\\
&= 3.2\times 10^{10} \bigg(\frac{M_{\odot}^2}{MM'}\bigg)\bigg(\frac{M_t}
{2M_{\odot}}\bigg)^{\!\!1/3}\!\! \bigg(\!\frac{\Omega}{0.1\,\textrm{s}^{-1}}
\bigg)^{\!\! -8/3}\,{\rm s},
\label{td}
\end{align}
where $M_t=M+M'$ is the total binary mass.
When a strong tidal resonance occurs, the orbital decay rate can be modified,
and we need to follow the evolution of the orbit and the mode amplitudes
simultaneously. The gravitational interaction energy between $M'$ and the modes
in star $M$ is
\begin{align}
W &= \int\! d^3x\,\, U(\br,t)\sum_\alpha a^\ast_\alpha(t)\, 
\delta\rho^\ast_\alpha(\br)\nonumber\\
&= -\sum_\alpha \frac{M'MR^2}{D^3}W_{lm}Q_\alpha
\,e^{-im\Phi}\,a^\ast_\alpha(t),
\end{align}
where we have restricted to the $l=2$ terms and set $G=1$. 
The orbital evolution equations, including the effects of
the modes, are then given by
\begin{align}
\ddot D-D\dot\Phi^2 =& -\frac{M_t}{D^2}-\sum_\alpha \frac{3M_t}{D^4}W_{lm}
Q_\alpha\,\, e^{im\Phi} a_\alpha  \nonumber\\
& -\frac{M_t}{D^2}\left(A_{5/2}+B_{5/2}\dot D\right),
\label{ddotd}\\
\ddot\Phi+\frac{2\dot D\dot\Phi}{D} =& \sum_\alpha
im \frac{M_t}{D^5}W_{lm}Q_\alpha\,\, e^{im\Phi} a_\alpha \nonumber\\
& -\frac{M_t}{D^2}B_{5/2}\dot\Phi.
\label{ddotphi}
\end{align}
The last terms on the right-hand side of equations (\ref{ddotd}) and (\ref{ddotphi}) are the leading-order gravitational
radiation reaction forces, with (see Lai \& Wiseman 1996 and references therein)
\begin{align}
&A_{5/2}=-\frac{8\mu}{5D}\dot D\left(18v^2+\frac{2M_t}{3D}-25\dot D^2\right),\\
&B_{5/2}=\frac{8\mu}{5D}\left(6v^2-\frac{2M_t}{D}-15\dot D^2\right),
\label{B}
\end{align}
where $\mu=MM'/M_t$ and $v^2=\dot D^2+(D\dot\Phi)^2$. In equations (\ref{ddotd})-(\ref{B}) we have set $G=c=1$.
We have dropped the other post-Newtonian terms since 
they have negligible effects on tidal excitations.
The mode amplitude equation is given by equation (\ref{eq:addot}), or,
\be
\ddot{b}_{\alpha} - 2im\Omega\dot{b}_{\alpha} + 
(\omega_{\alpha}^2-m^2\Omega^2-im\dot{\Omega})b_{\alpha} = 
\frac{M'W_{lm}Q_{\alpha}}{D^{l+1}},
\label{eq:bddot}
\ee
where 
\be 
b_\alpha=a_\alpha\, e^{im\Phi}.
\ee

\section{White Dwarf G-Modes and Tidal Coupling Coefficients}
\label{modes}

The non-radial adiabatic modes of a WD can be found by solving the
standard stellar oscillation equations, as given in, 
e.g., Unno et al.~(1989). 
The g-mode propagation zone in the star 
is determined by $\omega_\alpha^2<N^2$ amd $\omega_\alpha^2<L_l^2$, 
where $L_l=\sqrt{l(l+1)}{a_s/r}$ 
is the Lamb frequency ($a_s$ is the sound speed), and $N$ is the
Br\"unt-V\"ais\"al\"a frequency, as given by 
\be 
N^2 = g^2\bigg[\frac{d\rho}{d P} - \bigg(\frac{\partial\rho}{\partial P}
\bigg)_s\bigg]\ , 
\ee 
where $g$ the gravitational acceleration, and the
subscript \textquotedblleft$s$\textquotedblright \ means that the adiabatic derivative is taken. Alternatively,
$N^2$ can be obtained from (Brassard et al. 1991)
\begin{equation}
N^2 = \frac{\rho g^2 \chi_T}{P \chi_{\rho}} \Big( \nabla_s - \nabla + B\Big),
\end{equation}
where
\ba
&&\chi_T= \Big(\frac{\partial \textrm{ln P}}{\partial \textrm{ln} T}\Big)_{\rho,\{X_i\}},\quad 
\chi_\rho= \Big(\frac{\partial \textrm{ln P}}{\partial \textrm{ln} \rho}\Big)_{T,\{X_i\}},
\nonumber\\
&&\nabla= \frac{d\textrm{ln}T}{d\textrm{ln}P},\quad
\nabla_s = \Big(\frac{\partial \textrm{ln} T}{\partial \textrm{ln}P}\Big)_{s,\{X_i\}}.
\ea
The Ledoux term $B$ accounts for the buoyancy arising from composition gradient:
\begin{equation}
\label{ledoux}
B = -\frac{\chi_Y}{\chi_T}\frac{d \textrm{ln}Y}{d \textrm{ln}P},
\end{equation}
where
\begin{equation}
\chi_Y = \bigg(\frac{\partial \textrm{ln} P}{\partial \textrm{ln} Y} \bigg)_{\rho,T},
\end{equation}
and $Y$ is the mass fraction of helium.  This equation is valid for
a compositional transition zone containing helium
and one other element, as is the case for typical compositionally
stratified DA WD models. 

\begin{figure}
\begin{centering}
\includegraphics[scale=.35]{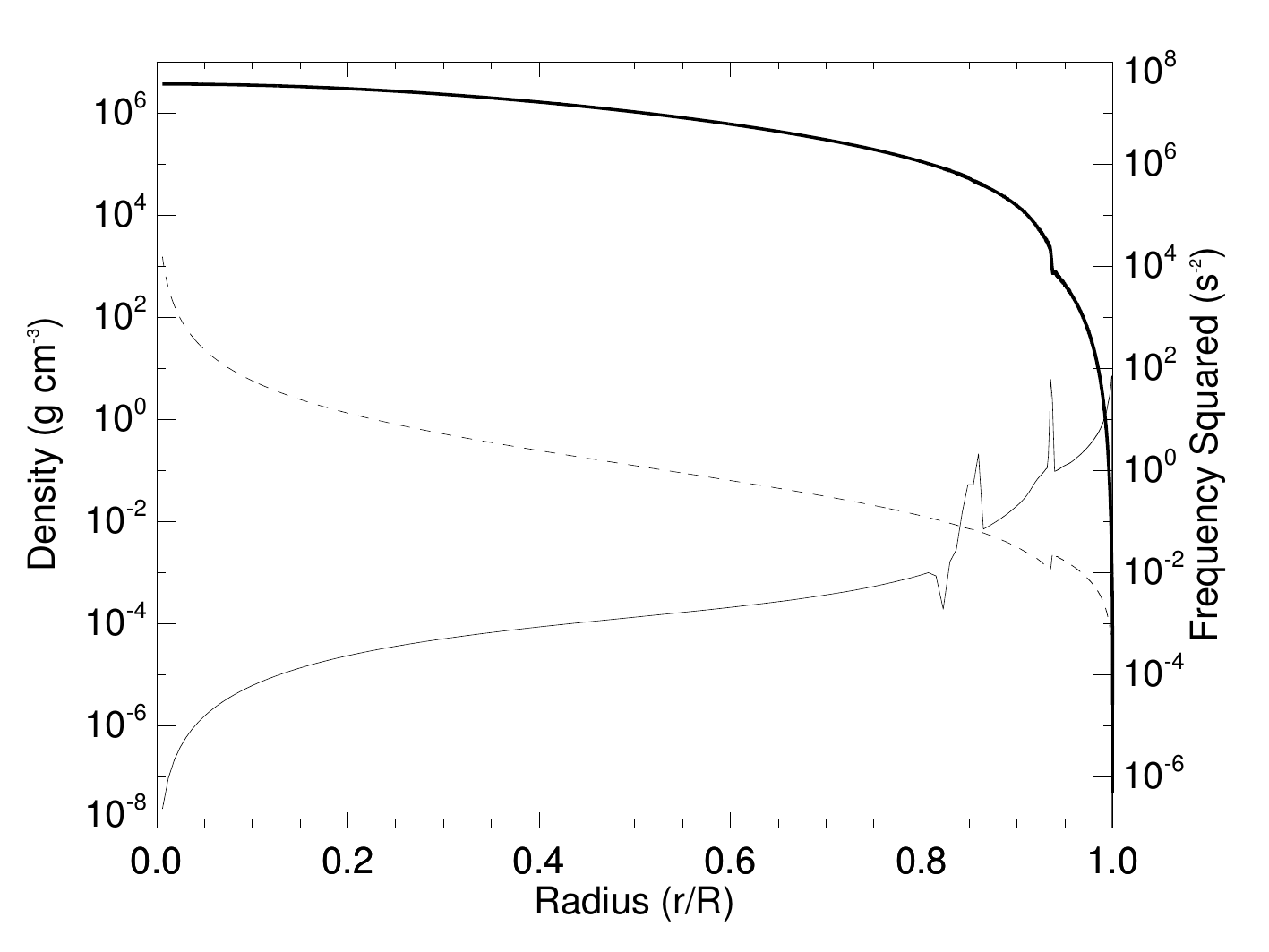}
\caption{\label{struc} The square of the Br\"unt V\"ais\"al\"a (solid line)
and Lamb (dotted line) frequencies and the density (thick solid line) 
as a function of normalized radius in a DA WD model, 
with $M=0.6M_\odot$, $R=8.97 \times 10^3$ km, $T_{\rm eff}= 10800$ K. The spikes in the Br\"unt
V\"ais\"al\"a frequency are caused by the composition changes from
carbon to helium, and from helium to hydrogen, respectively.}
\end{centering}
\end{figure}

Figure \ref{struc} shows the profiles of the Br\"unt Vais\"al\"a and
Lamb frequencies for one of the WD models adopted in this paper. These
models were provided by G. Fontaine (see Brassard 1991). Since the
pressure in the WD core is almost completely determined by electron
degeneracy pressure, $N^2 \propto \chi_{T}$ is very small except in
the non-degenerate outer layers.  As a result, g-modes are confined to
the outer layers of the star below the convection
zone. Lower-order modes have higher eigenfrequencies, so they are
confined to regions where $N^2$ is especially large, i.e., just below
the convection zone.  Higher order modes have lower eigenfrequencies
and can thus penetrate into deeper layers of the star where the value
of $N^2$ is smaller.
Cooler WDs have deeper convection zones that cause the modes to be confined to
deeper layers where $N^2$ is smaller. Consequently, the
eigenfrequencies and associated values of $Q_\alpha$ tend to be
smaller in cooler WDs due to the decreased value of $N^2$ in the
region of mode propagation.

\begin{table}
\centering
\caption{\label{table1} The eigenfrequency $\bomega$, tidal overlap parameter $\bQ$, 
and numerical f-mode overlap $c_0$ for the first 
six $l=2$ g-modes of a white dwarf model. The white dwarf model has $T_{\textrm{eff}} = 10800K$, 
$M=0.6 M_\odot$, and $R=8.97\times10^3$ km. 
Note that $\bomega$ and $\bQ$ are in dimensionless units such that $G=M=R=1$, 
and $(GM/R^3)^{1/2}/(2\pi)=0.053$~Hz.}
\begin{tabular}{@{}lccc}
n & $\bomega$ & $|\bQ|$ & $c_{\textrm{\tiny{0}}}$ \\
\hline
0 & 2.08 & 0.428 & 1 \\
\hline
1 & 0.298 & 1.27e-3 & -1.80e-6 \\
\hline
2 & 0.186 & 2.60e-3 & 3.91e-6 \\
\hline
3 & 0.125 & 3.25e-5 & 3.62e-8 \\
\hline
4 & 0.0900 & 8.91e-5 & 6.45e-7 \\
\hline
5 & 0.0821 & 4.24e-4 & 1.32e-6 \\
\hline
6 & 0.0715 & 6.91e-5 & -2.54e-6 \\
\end{tabular}
\end{table}

The other feature of WDs that strongly effects their g-modes is their
compositionally stratified layers. The sharp composition gradients
that occur at the carbon-helium transition and the helium-hydrogen
transition create large values of the Ledoux term $B$ [equation 
(\ref{ledoux})], resulting in sharp peaks in $N^2$ as seen in 
Figure \ref{struc}.  These peaks have a large effect on the WD g-modes,
leading to phenomena such as mode-trapping (e.g., Brassard 1991) and
irregular period spectra. Thus, the eigenfrequencies and
eigenfunctions of WD g-modes are very sensitive to WD models.

\begin{table}
\centering
\caption{\label{table2} Same as table 1, for a WD model of identical mass and composition 
but with $T_{\textrm{eff}} = 5080K$.}
\begin{tabular}{@{}lccc}
n & $\bomega$ & $|\bQ|$ & $c_{\textrm{\tiny{0}}}$ \\
\hline
0 & 2.01 & 0.439 & 1 \\
\hline
1 & 0.251 & 1.00e-3 & -2.03e-6 \\
\hline
2 & 0.156 & 2.40e-3 & -3.76e-6 \\
\hline
3 & 0.107 & 1.79e-5 & -1.25e-7 \\
\hline
4 & 0.0723 & 1.53e-5 & 4.95e-7 \\
\hline
5 & 0.0537 & 8.42e-5 & 1.52e-6 \\
\hline
6 & 0.0513 & 1.26e-4 & 1.42e-6 \\
\end{tabular}
\end{table}

Tables 1-2 give the $l=2$ f-mode and g-mode frequencies and their 
tidal coupling coefficients for two WD models. While the
full oscillation equations need to be solved to accurately determine the f-modes, the Cowling
approximation (in which the perturbation in the gravitational potential
is neglected) gives accurate results for g-modes. Since high-order g-modes
have rather small $|Q_\alpha|$, the mode eigenfunction must be solved
accurately to obtain reliable $Q_\alpha$. To 
ensure that this is achieved in our numerical integration, 
we use the orthogonality of the eigenfunctions to check the 
accuracy of the value of $Q_{\alpha}$ (see Reisengger 1994 for 
a study on the general property of $Q_\alpha$).
Since the numerical determination of an eigenfunction is
not perfect, it will contain traces of the other eigenfunctions, i.e.,
\be
(\bxi_\alpha)_{\rm num}=c_\alpha\bxi_\alpha+c_0\bxi_0+c_1\bxi_1
+\cdots,
\ee
with $c_{\alpha} \simeq  1$ and $|c_{\beta}| \ll 1$ for $\beta \neq \alpha$. 
This means that the numerical tidal overlap integral is
\begin{align}
(Q_{\alpha})_{\rm num} &= \langle \nabla (r^lY_{lm})| (\bxi_{\alpha})_{\rm num} \rangle \nonumber \\
&=  c_{\alpha}Q_{\alpha} + c_{0}Q_{0} + c_1 Q_1 + \dots
\label{numoverlap}
\end{align}
Since $|Q_0|$ (for the f-mode) is of order unity, while $|Q_\alpha|\ll 1$ for g-modes,
to ensure $(Q_\alpha)_{\rm num}$ accurately represents the actual $Q_\alpha$, we require
\be
|c_0| \simeq 
|\langle\bxi_0|\bxi_{\alpha} \rangle_{\rm num}|
\ll |Q_\alpha|.
\ee
The results shown in tables 1-2 reveal that $|c_0|$ is always more
than an order of magnitude less than $\bQ$, so the above condition
is satisfied for the modes computed in this paper.

\begin{figure}
\begin{centering}
\includegraphics[scale=.35]{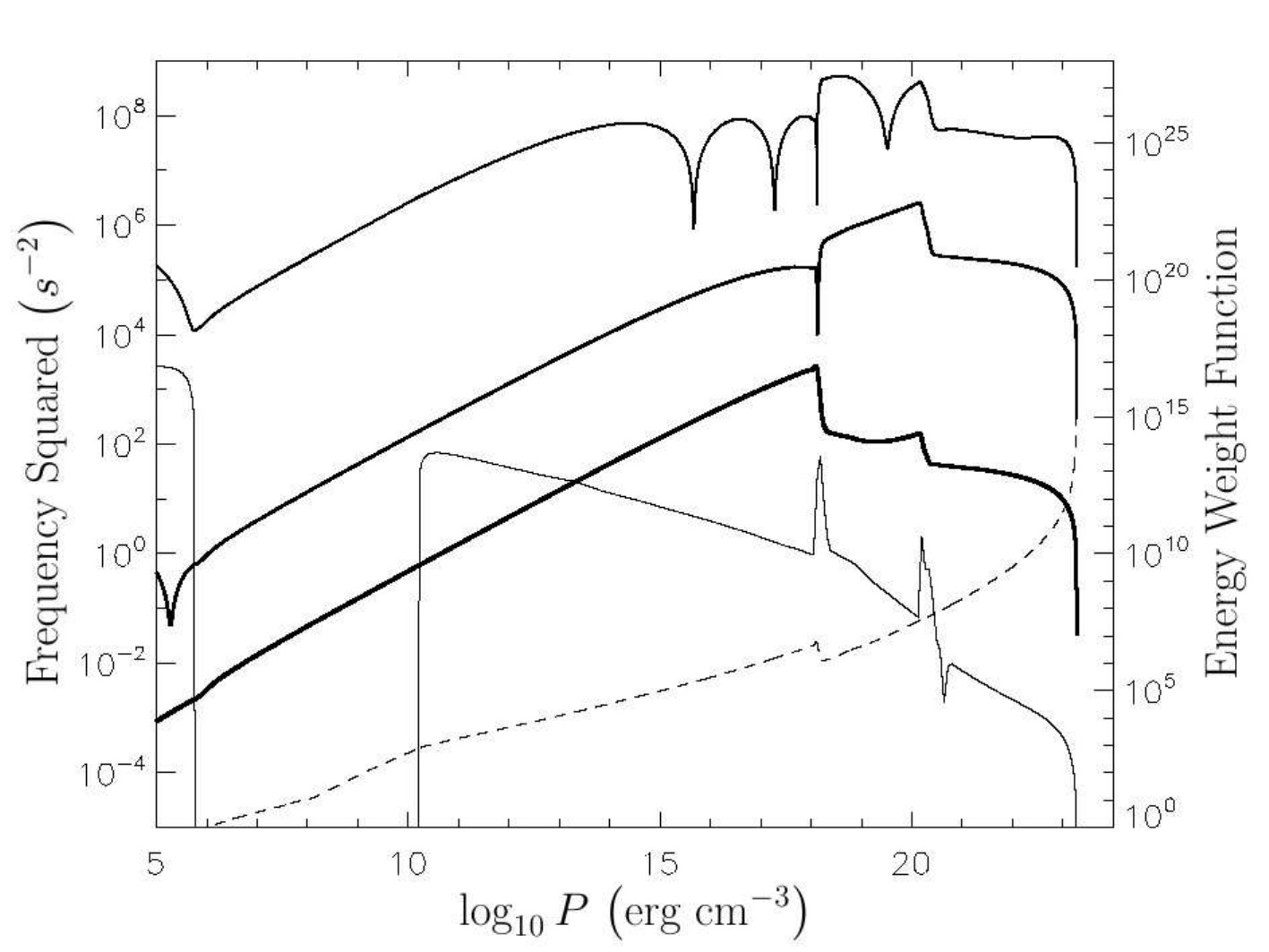}
\caption{\label{modestruc0090} The mode energy weight functions for the $n=1$ (thickest
line), $n=2$ (thick line), and $n=5$ (top line) modes (all for $l=2$)
for a WD with $T_{\textrm{eff}}=10800$K, 
$M=0.6 M_\odot$,$R= 8.97\times 10^3$km, displayed as a function of $\textrm{log}P$
so that the structure of the outer layers of the WD is more
evident. The y-axis for a given mode is intended only to
show the relative value of the weight function. The squares of the
Br\"unt V\"ais\"al\"a (thin solid line) and Lamb (dotted line)
frequencies are displayed to demonstrate how their values constrain
the region of mode propagation.}
\end{centering}
\end{figure}

\begin{figure}
\begin{centering}
\includegraphics[scale=.35]{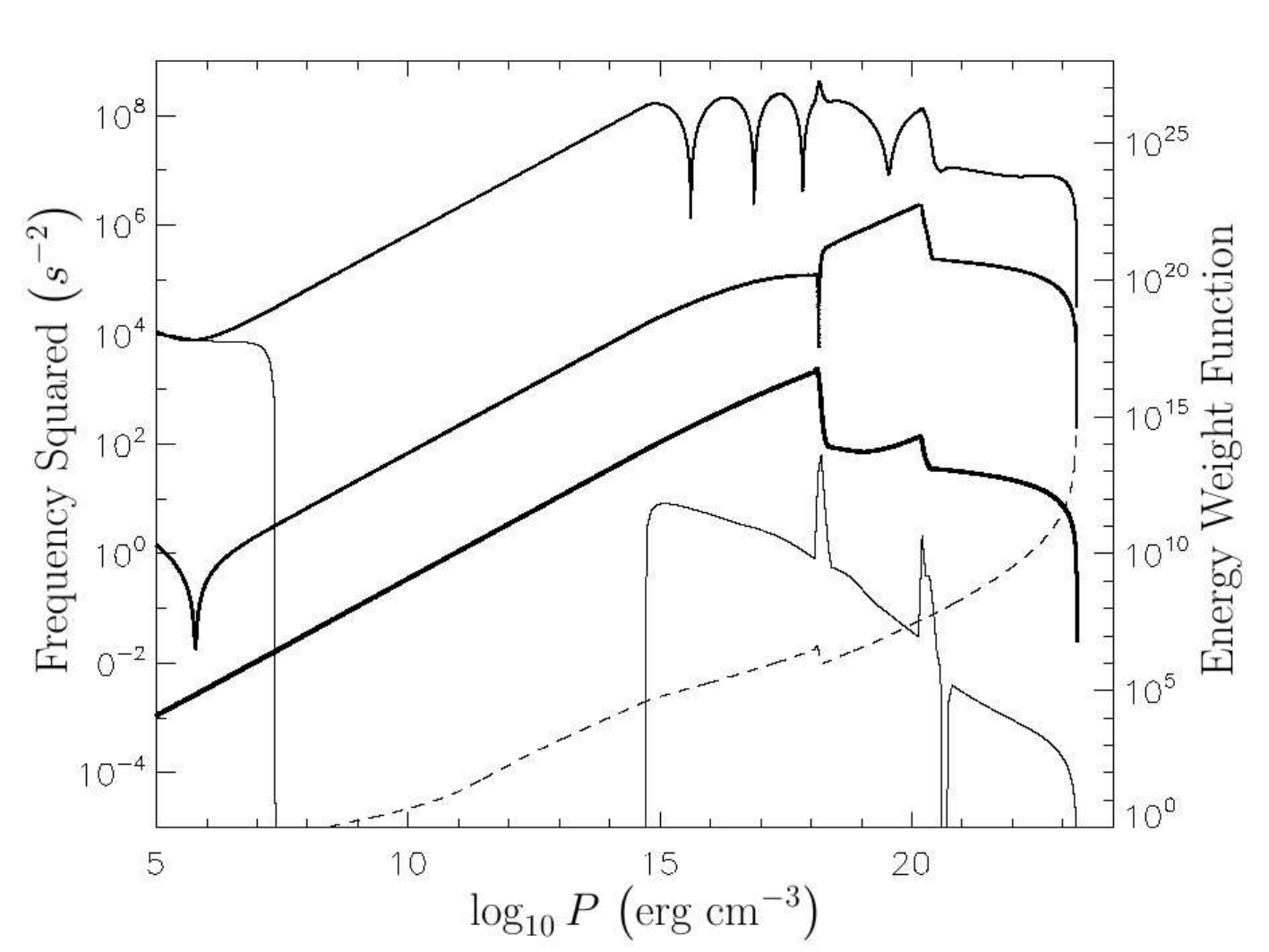} 
\caption{\label{modestruc0130} Same as Figure \ref{modestruc0090}, except for a WD model
  with $T_{\textrm{eff}}=5080$K. Note that the convection zone extends
  deeper in this model, pushing the modes to larger depths.}
\end{centering}
\end{figure}

We note from tables 1-2 that while in general higher-order g-modes
tend to have smaller $|Q_\alpha|$, the dependence of $|Q_\alpha|$ on
the mode index $n$ is not exactly monotonic. This is the result of the 
mode trapping phenomenon associated with composition discontinuities in the WD.
To see this, we note that a mode with amplitude $\bxi_\alpha$
has energy given by $E_\alpha=\omega_\alpha^2\int\! d^3x\,\rho\,|\bxi_\alpha|^2$,
thus we can define the mode energy weight function
\be
\frac{dE_\alpha}{d\ln P}=\omega_\alpha^2
\rho\, r^2 \big[\xi_r^2+l(l+1)\xi_{\perp}^2\big]H_p,
\ee
where $H_p=dr/d\ln P=P/(\rho g)$ is the pressure scale height, 
and we have used
\be
\bxi_\alpha=\left[\xi_r(r)\,{\bf e}_r+r\xi_\perp(r)\nabla\right]Y_{lm}
\ee
(${\bf e}_r$ is the unit vector in the $r$-direction).
Figures \ref{modestruc0090} and \ref{modestruc0130} display the weight
functions for several g-modes of WD models.  We can see that the weight
functions for all the low-order modes are largest in the region below
the convective zone near the spikes in $N^2$ produced by the
composition gradients.  For the modes shown in Figure
\ref{modestruc0090}, the smooth fall-off of the weight function just
below the convective zone indicates that these modes are confined by
the falling value of the Lamb frequency in this region. The weight
functions of higher-order modes and the modes in WDs with deeper
convective zones may drop sharply at the convective boundary,
indicating that these modes are trapped by the convective zone rather
than the decreasing Lamb frequency.

The weight functions also reveal the phenomenon known as mode trapping
caused by the composition gradients. Mode trapping is especially
evident for the $n=2$ mode, as it is confined to the helium layer
between the spikes in $N^2$.  It is clear that the mode is reflected
by the carbon-helium boundary at larger depths and by the
helium-hydrogen boundary at shallower depths. See Brassard (1991) for
a more detailed description of the effects of mode trapping.

The weight function is essentially the energy of a mode as a function
of radius, so it tells us where orbital energy is deposited when a
mode is excited. Since the weight function is largest in the hydrogen
and helium layers just below the convection zone, most of the mode
energy exists in this region of the WD. Thus, if the mode is damped,
most of the mode energy will be damped out in this region. 

\section{Numerical Results for Mode-Orbit Evolution Through Resonance}

Having obtained the mode frequency and the tidal coupling coefficient,
we can determine the combined evolution of the resonant mode and the
binary orbit using equations (\ref{eq:addot}), (\ref{ddotd}), and
(\ref{ddotphi}). These are integrated from well before resonance until
well after the resonance is complete. The initial mode amplitude
$b_\alpha$ and its derivative $\dot b_\alpha$ (prior to a resonance)
are obtained by dropping the $\ddot b_\alpha$, $\dot b_\alpha$ and
$\dot\Omega$ terms in equation (\ref{eq:bddot}), giving
\begin{align}
& b_{\alpha} \simeq \frac{M'W_{lm}Q_{\alpha}}{D^{l+1}(\omega_{\alpha}^2-m^2\Omega^2)},
\label{b0}\\
& \dot{b}_{\alpha} \simeq \bigg[-(l+1)\frac{\dot{D}}{D} + 
\frac{2m^2\Omega\dot{\Omega}}{\omega_{\alpha}^2 - m^2 \Omega^2} \bigg] b_{\alpha},
\label{b0dot}
\end{align}
with $\dot\Omega\simeq -(3\dot D/2D)\Omega$. These expressions are valid for
for $(\omega_\alpha/m\Omega)^2-1\gg \dot\Omega/(m\Omega^2)\simeq 3/(2m\Omega t_D)$
(see section 5).

The evolution equations (\ref{eq:addot}), (\ref{ddotd}), and (\ref{ddotphi}) form a very stiff set of differential
equations. The reason for this is that the problem involves two vastly
different time scales: the orbital decay time scale which is on the
order of thousands of years, and the orbital time scale (or the resonant 
mode oscillation period) which is on the order of minutes. Consequently, 
a typical Runge-Kutta scheme would require the integration of millions of
orbits, demanding a high degree of accuracy for each orbit. 
To avoid this problem, we employ the Rossenberg stiff equation
technique (Press et al.~2007). The integrator requires a Jacobian matrix of second derivatives, meaning that we
need to supply the $8\times 8$ matrix of second derivatives corresponding to
our 8 first-order differential equations. The evolution equations are sufficiently
simple that this matrix can be found analytically. 

\begin{figure}
\begin{centering}
\includegraphics[scale=.35]{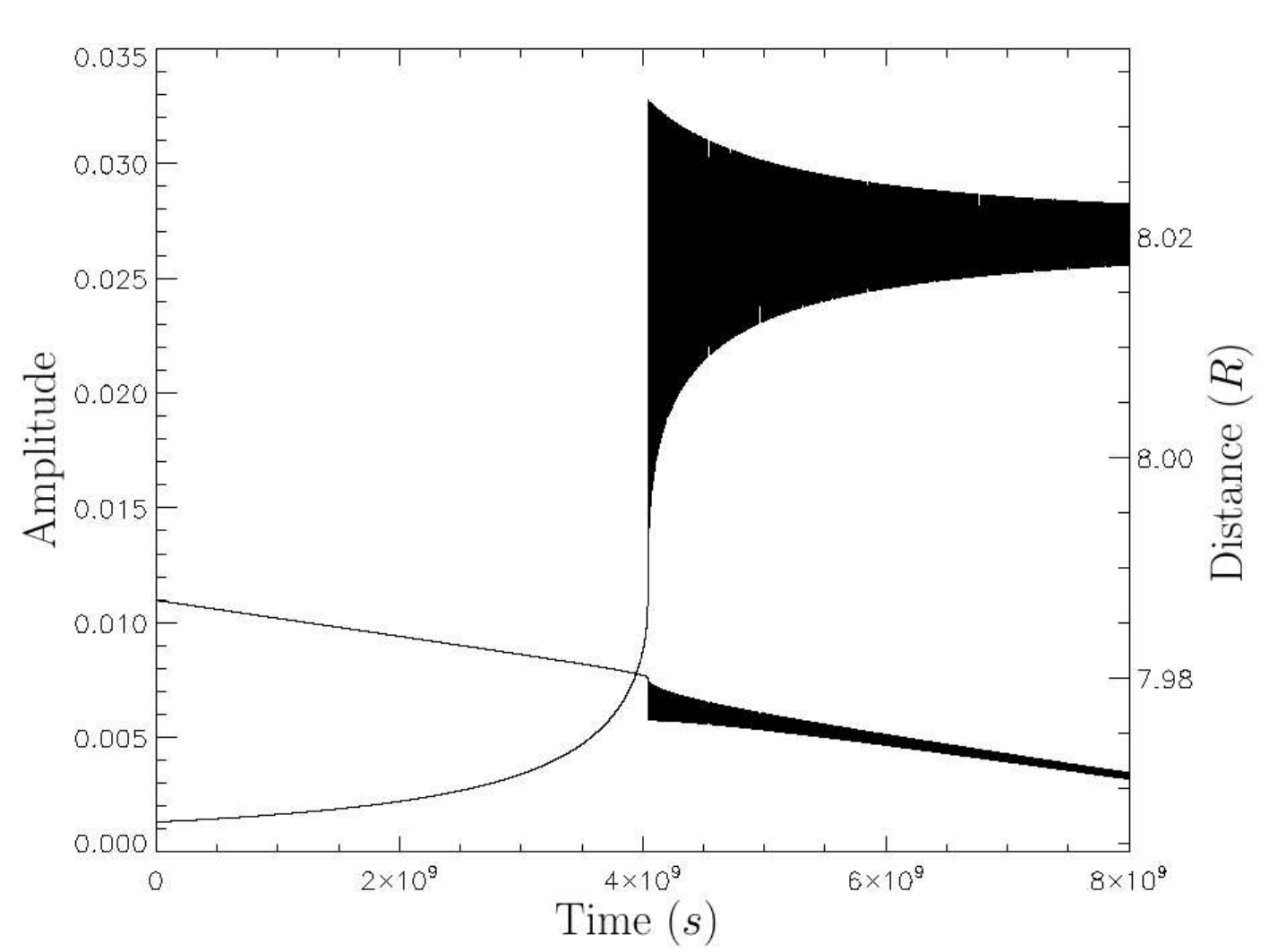}
\caption{\label{amp} Mode amplitude $|b_\alpha|$ and orbital distance $D$ as a
  function of time during resonance. The amplitude and distance
  oscillate after resonance, causing the curves to appear as filled
  shapes due to the short period of the oscillations with respect to
  the total integration time. These oscillations occur due to the
  continued interaction between the excited mode and the orbit. Note
  the sharp drop in orbital distance at resonance, which is caused by
  the transfer of orbital energy into the mode. The mode parameters
  are given in table I ($n=3$). The companion mass is $M'=M=0.6 M_\odot$. }
\end{centering}
\end{figure}

Figure \ref{amp} depicts an example of the mode amplitude and orbit
evolution near the resonance for the $n=3$ g-mode.  Before
the resonance, the mode must oscillate with the same frequency as the
binary companion, so the amplitude $|b_{\alpha}|$ is smooth. After
the resonance, the mode oscillates at its eigenfrequency
(which is now different from the forcing frequency). The
amplitude of the mode continues to fluctuate after resonance because
it is still being forced by the binary companion, although the
amplitude of these fluctuations diminishes over time as the orbit
moves further from resonance.

\begin{figure}
\begin{centering}
\includegraphics[scale=.35]{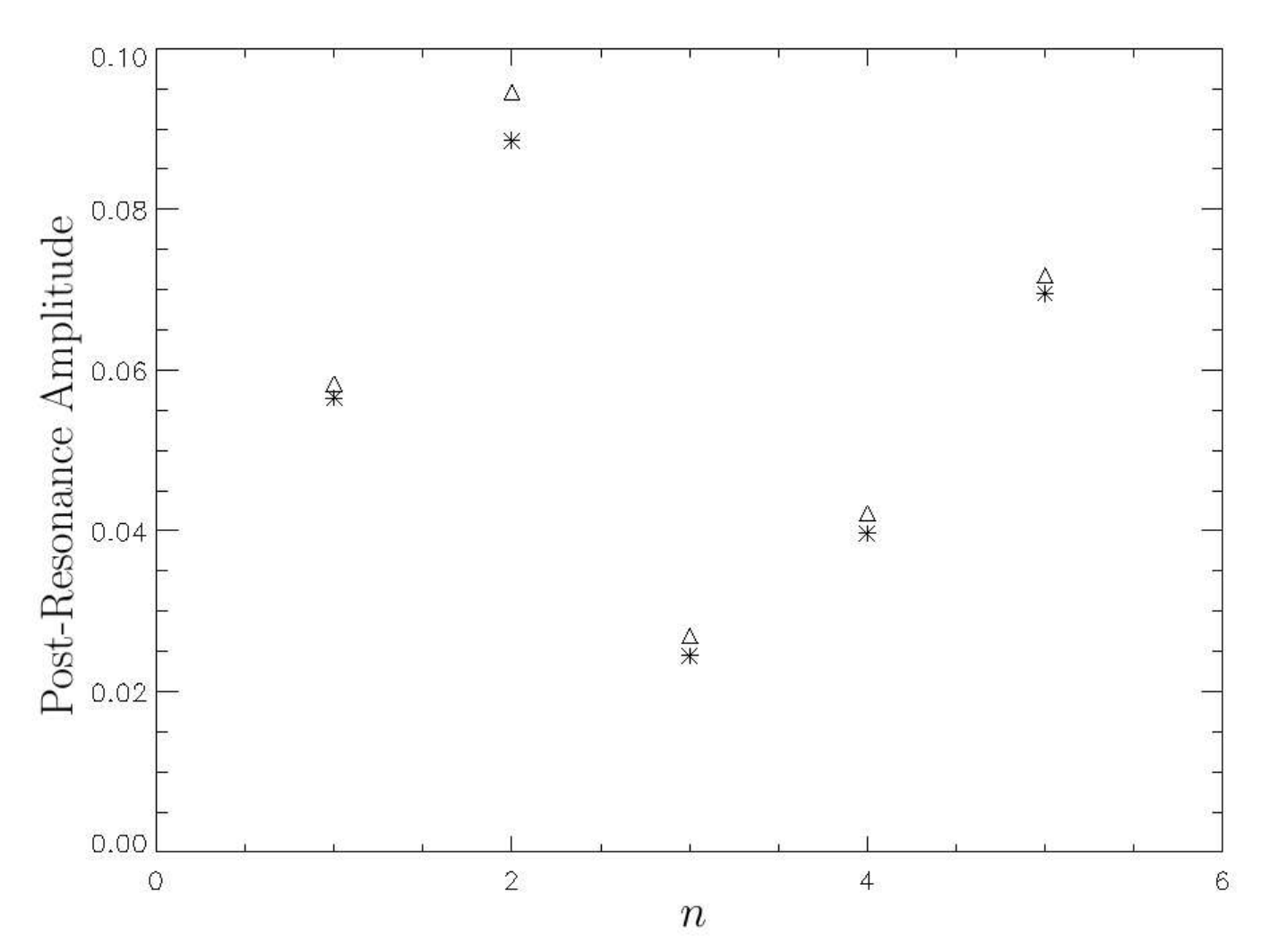}
\caption{\label{ampest} The average post-resonance amplitudes for the first five
  modes given in Table I. 
The open triangles mark the results obtained from numerical integration,
and the asterisks indicate the results predicted by the analytical 
estimate described in section \ref{estimate}.
The analytical estimates are usually accurate within a factor of 10\%. 
The $n=2$ and $n=5$ modes are trapped modes.} 
\end{centering}
\end{figure}

Figure \ref{ampest} displays the average post-resonance amplitudes for
the first five modes given in Table I.
Note that no mode exceeds a maximum amplitude
of $|b_{\alpha}|=0.1$, and we expect that our linear approximation 
is a reasonable first approach to the problem before non-linear effects 
can be included. In general, lower-order modes reach larger amplitudes 
(due to their larger coupling coefficients), but higher-order modes with an
abnormally high value of $Q_{\alpha}$ (trapped modes) may reach large
amplitudes as well.

\begin{figure}
\begin{centering}
\includegraphics[scale=.35]{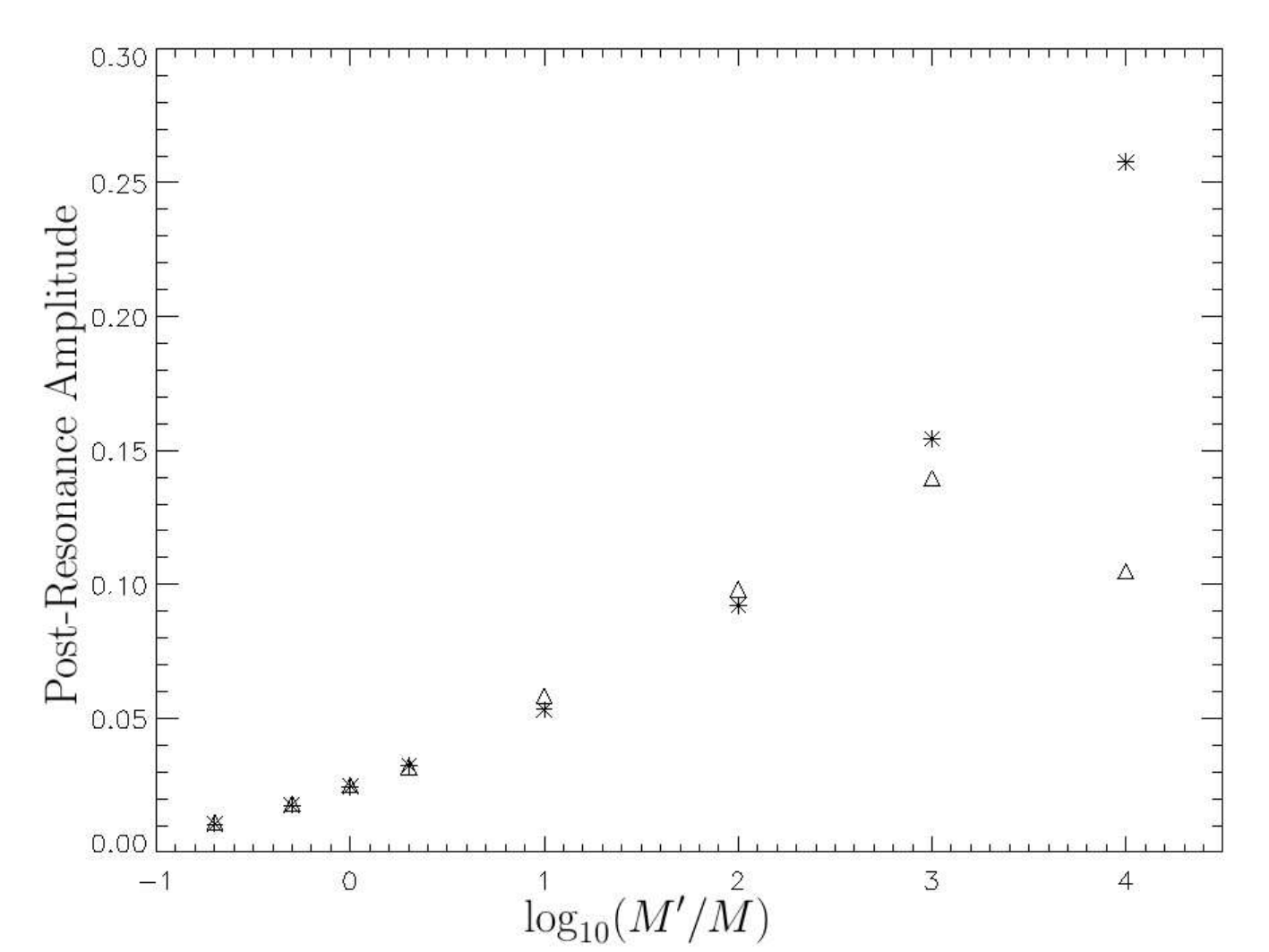}
\caption{\label{ampestmass} The average post-resonance amplitude $|b_\alpha|$ for the
  $n=4$ mode given in Table I as a function of the mass of the binary
  companion. The post-resonance amplitude increases with the mass of
  the binary companion as predicted by the analytical estimate except
  for very high companion masses.}
\end{centering}
\end{figure}

\section{Analytic Estimate of the Resonant Mode Amplitude}
\label{estimate}

Here we provide an analytic estimate of the mode amplitude attained
during a resonance as well as the temporal duration of the resonance
(i.e., the characteristic time during which the resonant mode receives
most of its energy from the orbit).

For a WD oscillation mode with frequency $\omega_\alpha$, the resonant orbital 
radius is 
\be
D_\alpha=\left(\frac{m^2M_t}{\omega_\alpha^2}\right)^{1/3}.
\ee
Prior to the resonance, as the orbital radius $D$ decreases,
the mode amplitude grows gradually according to equation (\ref{b0}). 
At the same time, the orbit also loses its energy to gravitational waves (GWs)
at the rate 
\be
\dot E_{\rm GW}=-\frac{32(MM')^2 M_t}{5c^5D^5}=-\frac{MM'/2D}{t_D},
\ee
where $t_D$ is given by equation (\ref{td}). We can define the beginning of 
resonance as the point where the orbital energy is transferred to the mode 
faster than it is radiated away by GWs. That is, the resonance begins at 
the radius $D=D_{\alpha+}(>D_\alpha)$, as determined by
\be
\dot{E}_{\alpha} = |\dot{E}_{\rm GW}|,
\ee
where $E_{\alpha}$ is the energy contained in the mode. Near the resonance, the mode oscillates at the frequency close to 
$\omega_\alpha$, so we can write the mode energy as 
$E_\alpha\simeq 2\omega_\alpha^2 b_\alpha^2$ (including
both the $m=2$ and $m=-2$ terms), assuming $\omega_\alpha\gg
|\dot b_\alpha/b_\alpha|$. Thus we have
$\dot E_\alpha\simeq 4 \omega_{\alpha}^2 b_{\alpha} \dot{b}_{\alpha}$.
Using equations (\ref{b0}) and (\ref{b0dot}) for $b_\alpha$ and $\dot b_\alpha$, we find
\be
\omega_\alpha^2-(m\Omega)^2 = \left[\frac{24\omega_\alpha^4M'(W_{lm}Q_\alpha)^2
}{MD^5}\right]^{1/3} {\rm at}~~D=D_{\alpha+}
\ee
or
\be
D_{\alpha+}=D_\alpha\left[1+0.436 \mratio^{\!\!1/3}
\!\!\!\mtratio^{\!\!-5/9}\!\!\!{\bar\omega}_\alpha^{4/9}
{\bar Q}_\alpha^{2/3}\right],
\label{dalpha}
\ee
where we have used $l=m=2$ and $\bomega$, $\bQ$ are in dimensionless
units where $G=M=R=1$. The mode energy at $D=D_{\alpha+}$ is
\begin{align}
E_\alpha(D_{\alpha+})=& 0.0701\!\mratio^{4/3}\!\mtratio^{-8/9} \nonumber \\
& \times \bomega^{10/9}(W_{lm}\bQ)^{2/3}\frac{M^2}{R}.
\label{eq:ealphad}
\end{align}
Since for $D<D_{\alpha+}$, the orbital energy will be deposited into the mode 
much faster than it is being radiated away, we approximate that 
all the orbital energy between $D_{\alpha+}$ and $D_\alpha$ is transferred 
to the mode. Thus the mode energy increases by the amount
\be
\Delta E_\alpha = 2\times 
\bigg(\frac{MM'}{2D_\alpha} - \frac{MM'}{2D_{\alpha+}}\bigg),
\ee
where we have multiplied by a factor of two to account for the fact that 
energy is also deposited at a nearly equal rate before 
resonance as it is after resonance. Using equation (\ref{dalpha}), we find
\begin{align}
\label{echange}
\Delta E_\alpha=& 0.2804\!\mratio^{4/3}\!\mtratio^{-8/9} \nonumber \\
& \times {\bar\omega}_\alpha^{10/9}(W_{lm}{\bar Q}_\alpha)^{2/3}\frac{M^2}{R}.
\end{align}
This is exactly four times of equation (\ref{eq:ealphad}). Thus the maximum mode
energy after resonance is $E_{\alpha,\max}=E_\alpha(D_{\alpha+})+
\Delta E_\alpha$, or
\ba
&&E_{\alpha,\max}\simeq 5.75\times 10^{-4}
\!\mratio^{4/3}\!\!\mtratio^{-8/9}\nonumber\\
&&\qquad\quad\times \left(\frac{\bomega}{0.2}\right)^{10/9}\!\!
\left(\frac{\bQ}{10^{-3}}\right)^{2/3}\!\!\frac{M^2}{R}.
\label{eq:emax}
\ea
The corresponding maximum mode amplitude is 
\begin{align}
b_{\alpha,\max}\simeq & 8.48\times 10^{-2}
\!\mratio^{2/3}\!\!\mtratio^{-4/9} \nonumber \\
& \times \left(\frac{\bomega}{0.2}\right)^{-4/9}\!\!
\left(\!\frac{\bQ}{10^{-3}}\!\right)^{1/3}.
\label{eq:bmax}
\end{align}

Figures \ref{ampest} and \ref{ampestmass} compare our numerical
results with the analytical expressions
(\ref{eq:emax})-(\ref{eq:bmax}). We find good agreement for all the WD
resonant modes considered. Figure \ref{ampest} verifies the dependence
of $b_{\alpha,\max}$ on the mode frequency and the value of $\bQ$,
while figure \ref{ampestmass} verifies the dependence of
$b_{\alpha,\max}$ on the mass of the binary companion (except for the highest mass cases discussed below). Therefore
equations (\ref{eq:emax}) and (\ref{eq:bmax}) provide fairly accurate
estimates of the mode amplitude and energy without performing
numerical integrations.

For the very high companion masses $(M' \go 10^3 M)$ shown in Figure
\ref{ampestmass}, our analytical formula significantly overestimates
the post-resonance amplitude.  The reason for this is that the
gravitational decay time scale is shorter if the companion is more
massive. If the companion is massive enough, the orbit will decay
through resonance due to gravitational radiation before the orbital
energy of equation (\ref{echange}) can be deposited in the mode (see
below). Consequently, the amplitude to which a mode is excited
decreases if the mass of the companion becomes very high.
Therefore our analytical formula
overestimates the post-resonance amplitude for extremely massive
companions. For any reasonable WD or NS masses, our analytical
estimate is accurate, but for a super-massive black hole the estimate
may become inaccurate.

It is interesting that the above analytical results for the resonant
mode energy is independent of the gravitational wave damping time scale
$t_D$, in contrast to the NS/NS or NS/BH binary cases. In fact, for the
above derivation to be valid, the following four conditions must be
satisfied at $D_{\alpha+}$:
\begin{align}
{\rm (i)}\quad & \omega_\alpha\gg 
\left|\dot b_\alpha/b_\alpha\right|,\\
{\rm (ii)}\quad & \omega_\alpha^2-(m\Omega)^2\gg m\dot\Omega,\\
{\rm (iii)}\quad & \omega_\alpha^2-(m\Omega)^2\gg 2m\Omega
\left|\dot b_\alpha/b_\alpha\right|,\\
{\rm (iv)}\quad & \omega_\alpha^2-(m\Omega)^2\gg 
\left|\ddot b_\alpha/b_\alpha\right|.
\end{align}
Conditions (i) and (ii) both lead to
\be
\label{i}
\omega_\alpha t_D\gg 3\left[\frac{\omega_\alpha^2}{\omega_\alpha^2
-(m\Omega)^2}\right];
\ee
condition (iii) gives
\be
\label{ii}
\omega_\alpha t_D\gg 6\left[\frac{\omega_\alpha^2}{\omega_\alpha^2
-(m\Omega)^2}\right]^2;
\ee
and condition (iv) yields
\be
\label{iii}
\omega_\alpha t_D\gg \sqrt{18}\left[\frac{\omega_\alpha^2}{\omega_\alpha^2
-(m\Omega)^2}\right]^{3/2}.
\ee
In equations (\ref{i})-(\ref{iii}), the right-hand sides should be evaluated at $D_{\alpha+}$.
Clearly, condition (iii) is most constraining. With 
\begin{align}
\label{constrain}
&\frac{\omega_\alpha^2-(m\Omega)^2}{\omega_\alpha^2}=3\left(\frac{D_{\alpha+}
-D_\alpha}{D_\alpha}\right)\nonumber\\
&\quad = 0.0064\!\mratio^{\!\!1/3}\!\!\mtratio^{\!\!-5/9}\!\!
\left(\!\frac{\bomega}{0.2}\!\right)^{\!\!4/9}
\!\!\left(\!\frac{\bQ}{10^{-3}}\!\right)^{\!\!2/3},
\end{align}
we see that condition (iii) is satisfied if 
\begin{align}
\label{tdgg}
t_D \gg & 
7.3\times 10^5\left({R^3\over GM}\right)^{1/2}
\mratio^{\!\!-2/3}\!\!\! \mtratio^{\!\!10/9}\!\!\! \nonumber\\ 
&\times \left(\!\frac{\bomega}{0.2}\!\right)^{\!\!-17/9}
\!\!\left(\!\frac{\bQ}{10^{-3}}\!\right)^{\!\!-4/3}\!\!.
\end{align}
Since $t_D$ is on the order of a thousand years or more for orbital
frequencies comparable to WD g-modes, condition (iii) is always satisfied for 
WD/WD or WD/NS binaries. On the other hand, the conditions (i)-(iv) are not
all satisfied for NS/NS or NS/BH binaries.

For very massive companions, the inequality of equation (\ref{tdgg}) may
not hold. Using equation (\ref{td}) for $t_D$, equation (\ref{tdgg})
implies (for $M_t/M\simeq M'/M$)
\begin{align}
\label{mll}
{M'\over M} \ll & 5.5\times10^4 \left({\bomega\over 0.2}\right)^{\!\!-7/10}\!\!
\left({\bQ\over 10^{-3}}\right)^{\!\!6/5}\!\! \nonumber \\
& \times \left(\frac{M}{M_\odot}\right)^{\!\!-9/4} \!\!\left(\frac{R}{10^4{\rm km}}\right)^{\!\!9/4}.
\end{align}
The above inequality implies our estimates are valid for any feasible
companion except a super-massive black hole.\footnotemark We can also use this
inequality to examine the inaccuracy of our estimate in the highest
mass cases of Figure \ref{ampestmass}. Figure \ref{ampestmass} was
generated using the $n=3$ mode parameters listed in Table 1 for a WD
of $M=0.6M_\odot$ and $R=8.97\times10^3$ km. Plugging in these
parameters, equation (\ref{mll}) requires
\be
\frac{M'}{M} \ll 1.7\times10^3
\ee
for our analytical estimates in Figure \ref{ampestmass} to be
accurate. This explains why the analytical estimates of Figure
\ref{ampestmass} are accurate when $M' \lo 1000 M$ but diverge from the
numerical results when $M' \go 1000 M$.

\footnotetext{Obviously, our estimate would not apply for a WD in a highly eccentric orbit around an intermediate mass black hole, which may form in dense clusters as described by Ivanov \& Papaloizou (2007).}

Given the maximum mode amplitude reached during a
resonance, we can now estimate the temporal duration of the resonance. 
Letting $a_{\alpha} = c_{\alpha}e^{-i \omega_{\alpha} t}$, the mode amplitude
evolution equation (\ref{eq:addot}) becomes
\be
\ddot{c}_{\alpha} - 2 i \omega_{\alpha} \dot{c_{\alpha}} = 
\frac{M'W_{lm}Q_{\alpha}}{D^3}\,e^{i\omega_\alpha t-im\Phi}.
\label{eq:cddot}\ee
Assuming that during the resonance, $\omega_\alpha-m\Omega\simeq 0$, the right-hand-side
of equation (\ref{eq:cddot}) can be taken as a constant, we then have
\be
\dot c_{\alpha} \simeq \frac{iM'W_{lm}Q_{\alpha}}{2\omega_{\alpha}D^3},
\ee
i.e., the mode amplitude grows linearly in time. Thus, the duration of
the resonance is of order
\begin{align}
\label{tres}
t_{\rm res} &= \left|\frac{b_{\alpha,\max}}{\dot{c}_\alpha}\right|\nonumber\\
&\quad\simeq 3.42\,\mratio^{\!\!-1/3}\!\!\mtratio^{\!\!5/9}
\!\!\bomega^{-13/9}\bQ^{-2/3}\left(\frac{R^3}{M}\right)^{\!\!1/2}\nonumber\\
&\quad =3.50\times 10^3
\mratio^{\!\!-1/3}\!\!\mtratio^{\!\!5/9}\nonumber\\
&\qquad\times 
\left(\!\frac{\bomega}{0.2}\!\right)^{\!\!-13/9}\!\!
\left(\!\frac{\bQ}{10^{-3}}\!\right)^{\!\!-2/3}\left(\frac{R^3}{M}\right)^{\!\!1/2}.
\end{align}
Since the dynamical time $(R^3/GM)^{1/2}$
for typical WDs is on the order of one
second, the resonance duration is typically an hour or longer.  Note
that the above estimate is formally valid only when
$\left[\omega-m\Omega(D_{\alpha+})\right]t_{\rm res}\ll 1$, so that we
can set $\omega_\alpha - m \Omega \approx 0$ for the duration of the
resonance. Using equations (\ref{constrain}) and (\ref{tres}), we find
$[\omega_\alpha-m\Omega(D_{\alpha+})]t_{\rm res} \approx 1$ for
typical parameters. Thus we should consider equation (\ref{tres}) as
an order-of-magnitude estimate only.
Also, we can check that the GW energy
loss during the resonance, $\Delta E_{\rm gw}\simeq
(MM'/2D_\alpha)(t_{\rm res}/t_D)$, is much less than
$E_{\alpha,\max}$, justifying our derivation of $E_{\alpha,\max}$
given by equation (\ref{eq:emax}). 
Indeed, the above condition simplifies to equation \ref{mll}, since in
both cases it is the energy carried away by gravitational waves that
is limiting the mode growth.

We can use the same method to solve for the size of the
fluctuations in mode amplitude after resonance.  Due to the symmetry
of the harmonic oscillator, 
the fluctuation in mode amplitude about the mean value
after the resonance is identical to the zeroth-order estimate of the mode
amplitude before resonance 
[see eq.~(\ref{b0})], i.e.,
\begin{equation}
\Delta a \approx \frac{M'W_{lm}Q_{\alpha}}{D^{l+1}(m^2\Omega^2-\omega_{\alpha}^2)}.
\end{equation} 
These fluctuations occur with frequency
$m\Omega-\omega_{\alpha}$, since this is the difference in frequency
between the eigenfrequency at which the WD is oscillating and the
orbital forcing frequency. So, as the orbital frequency continues to
increase after the resonance, the amplitude of the fluctuations becomes
smaller while the frequency of the amplitude oscillations becomes higher.

\section{Effect of Mode Damping}

\begin{figure}
\begin{centering}
\includegraphics[scale=.35]{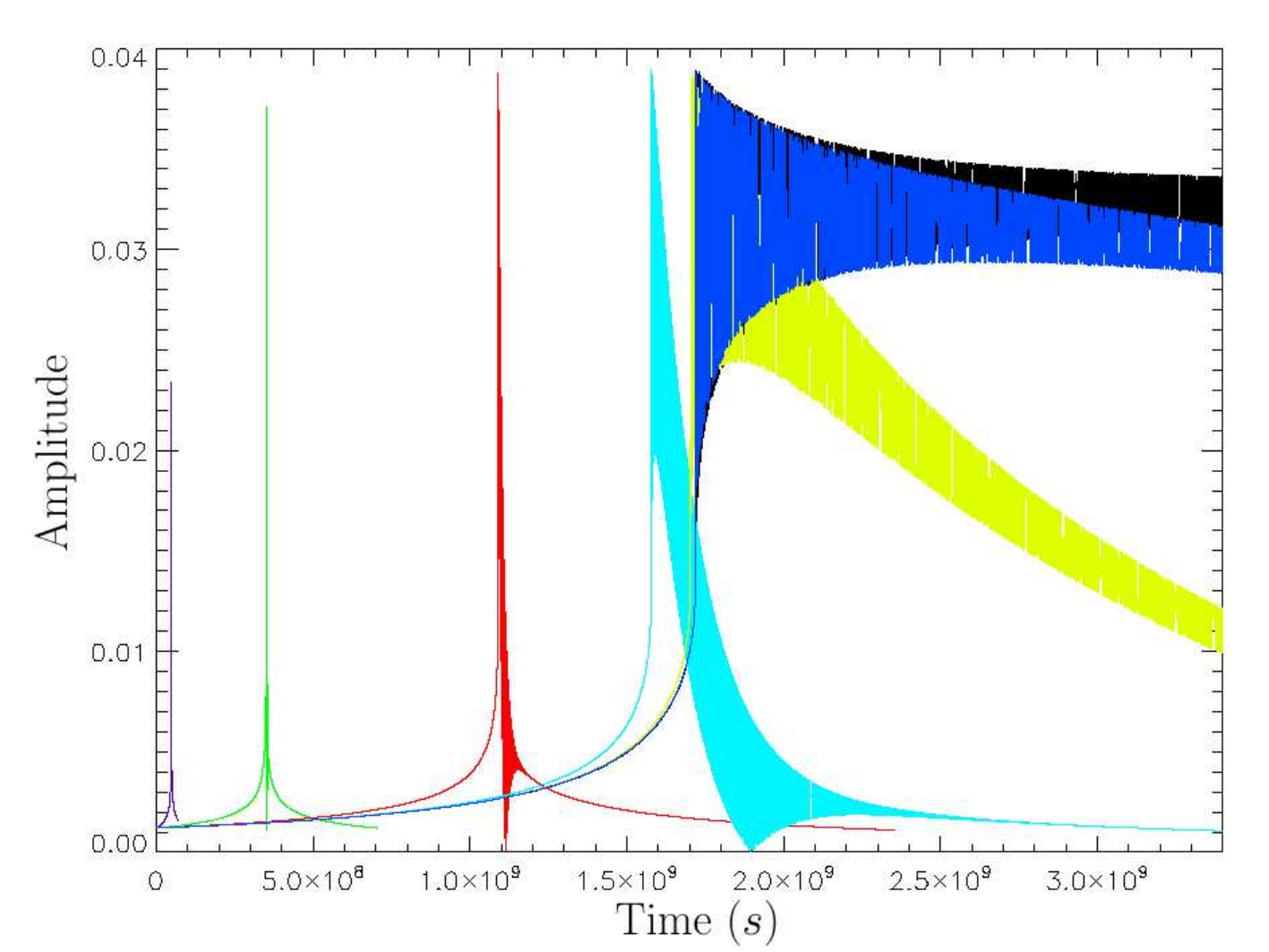}
\caption{\label{dampamp} The amplitude of a mode as a function of time near its resonance for different values of the damping coefficient $\gamma_\alpha$. The curves have $\gamma_\alpha = 0$ (black), $\gamma_\alpha = 10^{-9}$ (dark blue), $\gamma_\alpha = 10^{-8}$ (yellow-green), $\gamma_\alpha = 10^{-7}$ (light blue), $\gamma_\alpha = 10^{-6}$ (red), $\gamma_\alpha = 10^{-5}$ (green), and $\gamma_\alpha = 10^{-4}$ (purple). For this mode, we have set $\omega_\alpha = 0.1$s$^{-1}$ and $Q_\alpha = 1 \times 10^{-4}$ so that $t_{\rm res} \approx 10^5$s. Note that the damping term does not greatly affect the maximum mode amplitude except when $\gamma_\alpha \go 10^{-4}$, or when $\gamma_\alpha \omega_\alpha t_{\rm res} \go 1$. Also note that modes with larger values of $\gamma_\alpha$ evolve on a much shorter time scale because their orbits decay quickly due to the conversion of orbital energy into heat via mode damping.}
\end{centering}
\end{figure}

The results in the previous two sections neglect mode energy damping
in the WD. Since the duration of the resonance is much longer than the
mode period [see equation (\ref{tres})], internal mode damping could
affect the energy transfer during resonance if the damping rate is
sufficiently large. To address this issue, we incorporate a phenomenological damping rate
$-\gamma_\alpha \omega_{\alpha} \dot a_\alpha$ to the mode equation
(\ref{eq:addot}) to study how mode damping affects energy transfer during a resonance. Figure \ref{dampamp} shows
the excitation of a mode through resonance for different
values of $\gamma_\alpha$.  We see that, as expected, when the
internal damping time is larger than the resonance duration (equation
\ref{tres}), the maximum mode energy achieved in a resonance is
unaffected.

G-modes in white dwarfs are damped primarily by radiative diffusion.
For sufficiently large mode amplitudes, non-linear damping is also important
(e.g., Kumar \& Goodman 1996; see section \ref{discussion} for more discussion on this issue).
Wu (1998) presents estimates for the non-adiabatic radiative damping rates of WD
g-modes in terms of $\omega_i = \gamma \omega_r$. Extrapolating
Wu's values to $l=2$ modes for a white dwarf of temperature
$T = 10800$~K, we find $\gamma \sim 10^{-11}$ for modes near $n=1$ and $\gamma   
\sim 10^{-4}$ for high-order modes with $n \go 20$.
So, while the maximum amplitude of low-order modes is completely unaffected by
non-adiabatic effects, high-order modes will damp on time scales
similar to the excitation time scale. Therefore these high-order modes
will attain amplitudes somewhat smaller than estimated in the previous section.
This continual process of high-order mode damping may extract energy
out of the orbit more efficiently than discrete resonance events,
causing a steady decay of the binary's orbit.

\section{Discussion}
\label{discussion}

We have shown that during the orbital decay of compact white dwarf
binaries (WD/WD, WD/NS or WD/BH), a series of g-modes can be tidally
excited to large amplitudes (up to $0.1$ in dimensionless units) as
the orbital frequency sweeps through the resonant mode
frequencies. Such mode excitations can significantly affect the orbital
decay rate near resonance. Indeed, to properly calculate the resonant mode amplitude,
it is necessary to take into account of the back-reaction of the
excited modes on the orbit. One consequence of the resonant mode
excitations is that the low-frequency ($\lo 10^{-2}$~Hz) gravitational
waveforms emitted by the binary, detectable by LISA, will deviate
significantly from the point-mass binary prediction. This is in
contrast to the case of neutron star binaries (NS/NS or NS/BH) studied
previously (Reisenegger \& Goldreich 1994; Lai 1994; Shibata 1994; Ho
\& Lai 1999; Lai \& Wu 2006; Flanagan \& Racine 2006), where the
resonant mode amplitude is normally too small to affect the binary
orbital decay rate and the gravitational waveforms to be detected by
ground-based gravitational wave detectors such as LIGO and VIRGO.

In the case of WD binaries studied in this paper, the number of of orbits skipped as a result of a resonant mode excitation is
\be
\label{norbit}
\Delta N_{\textrm{orb}} = \frac{t_D}{P_{\textrm{orb}}} \frac{E_{\alpha,\textrm{max}}}{E_{\textrm{orb}}},
\ee
where $t_D$ is the gravitational wave decay time scale given in equation (\ref{td}), $P_{\textrm{orb}}$ is the orbital period, and $E_{\textrm{orb}}$ is the orbital energy at resonance. Using equation (\ref{eq:emax}) for $E_{\alpha,\textrm{max}}$, we find
\begin{align}
\label{norbit2}
\Delta N_{\textrm{orb}} = & 3.4\times10^6 \bigg(\!\frac{M_\odot^{17}}{M^{9}M'^{6}M_t^2}\!\bigg)^{\!\! 1/9} \nonumber \\
& \bigg(\!\frac{R}{10^4 \textrm{km}}\!\bigg)^{\!\! 2/3} \bigg(\!\frac{\bar{Q_\alpha}}{10^{-3}}\!\bigg)^{\!\!2/3} \bigg(\!\frac{\Omega}{0.1 \textrm{s}^{-1}}\!\bigg)^{\!\!-11/9}.
\end{align}
The number of skipped orbital cycles should be compared to the number of orbits in a decay time, expressed by
\begin{align}
\label{norbite}
\frac{d N_{\textrm{orb}}}{d \ln \Omega} & = \frac{1}{3\pi}\Omega t_D \nonumber \\
& = 3.3\times 10^{8} \bigg(\frac{M_{\odot}^2}{MM'}\bigg)\bigg(\frac{M_t}
{2M_{\odot}}\bigg)^{\!\!1/3}\!\! \bigg(\!\frac{\Omega}{0.1\,\textrm{s}^{-1}}
\bigg)^{\!\! -5/3}.
\end{align}
The huge number of skipped orbital cycles implies that such a resonant interaction would be important, but because the number of skipped orbital cycles is much smaller than the number of orbital cycles in a decay time, resonances will not dominate the decay process.

A second possible consequence of resonant mode excitations is that 
the large mode energy may lead to significant heating of the white dwarf
prior to the binary merger. Indeed, equation (\ref{eq:emax}) shows that 
for typical binary parameters, the mode energy can be a significant fraction 
($\sim 10^{-4}$--$10^{-3}$) of the gravitational binding energy of the star,
and comparable to the thermal energy. Indeed, the thermal energy of the WD is of order $E_{\textrm{th}} \approx \frac{M k T_c}{A m_p}$, where $T_c$ is the core temperature of the WD and $A$ is the mean atomic weight.  The ratio of post-resonance mode energy to thermal energy is then
\begin{align}
\frac{E_{\alpha,\textrm{max}}}{E_\textrm{th}} & \approx 1.7 \!\mratio^{\!\!4/3}\!\!\mtratio^{\!\!-8/9}\!\! \left(\!\frac{\bomega}{0.2}\!\right)^{\!\!10/9}\!\! \nonumber \\
& \times \left(\!\frac{\bQ}{10^{-3}}\!\right)^{\!\!2/3} \!\!\left(\!\frac{10^7 K}{T_c} \!\right) \left(\!\frac{M}{M_\odot}\!\right) \left(\!\frac{10^4 \textrm{km}}{R}\!\right).
\end{align}
This implies that the white dwarf may become bright thousands of years before binary merger.

A third consequence that may result from a resonance is significant spin-up of the WD. If we assume that all the angular momentum transferred to the WD during a resonance eventually manifests as rigid body rotation of the WD, the change in spin frequency of the WD is
\be
\label{omegas}
\Delta \Omega_s = \frac{E_{\alpha,\textrm{max}}}{I \Omega}, 
\ee
where $\Omega_s$ is the spin frequency of the WD and $I$ is its moment of inertia. Plugging in our expression for $E_{\alpha,\textrm{max}}$, we find
\begin{align}
\label{omegas2}
\Delta \Omega_s = & 0.29 \left(\frac{0.2}{\kappa}\right)
\!\mratio^{\!\!4/3}\!\!\mtratio^{\!\!-8/9}\nonumber \\
&\qquad\quad\times \left(\frac{\bomega}{0.2}\right)^{\!\!-8/9}\!\!
\left(\frac{\bQ}{10^{-3}}\right)^{\!\!2/3}\!\! \Omega,
\end{align}
where $\kappa = I/(MR^2) \approx 0.2$. We can thus see that a given resonance may deposit enough angular momentum to completely spin up the WD (or significantly alter its spin) by the time the mode damps out. This implies that mode resonances are potentially very important in the spin synchronization process.

\begin{figure}
\begin{centering}
\includegraphics[scale=.35]{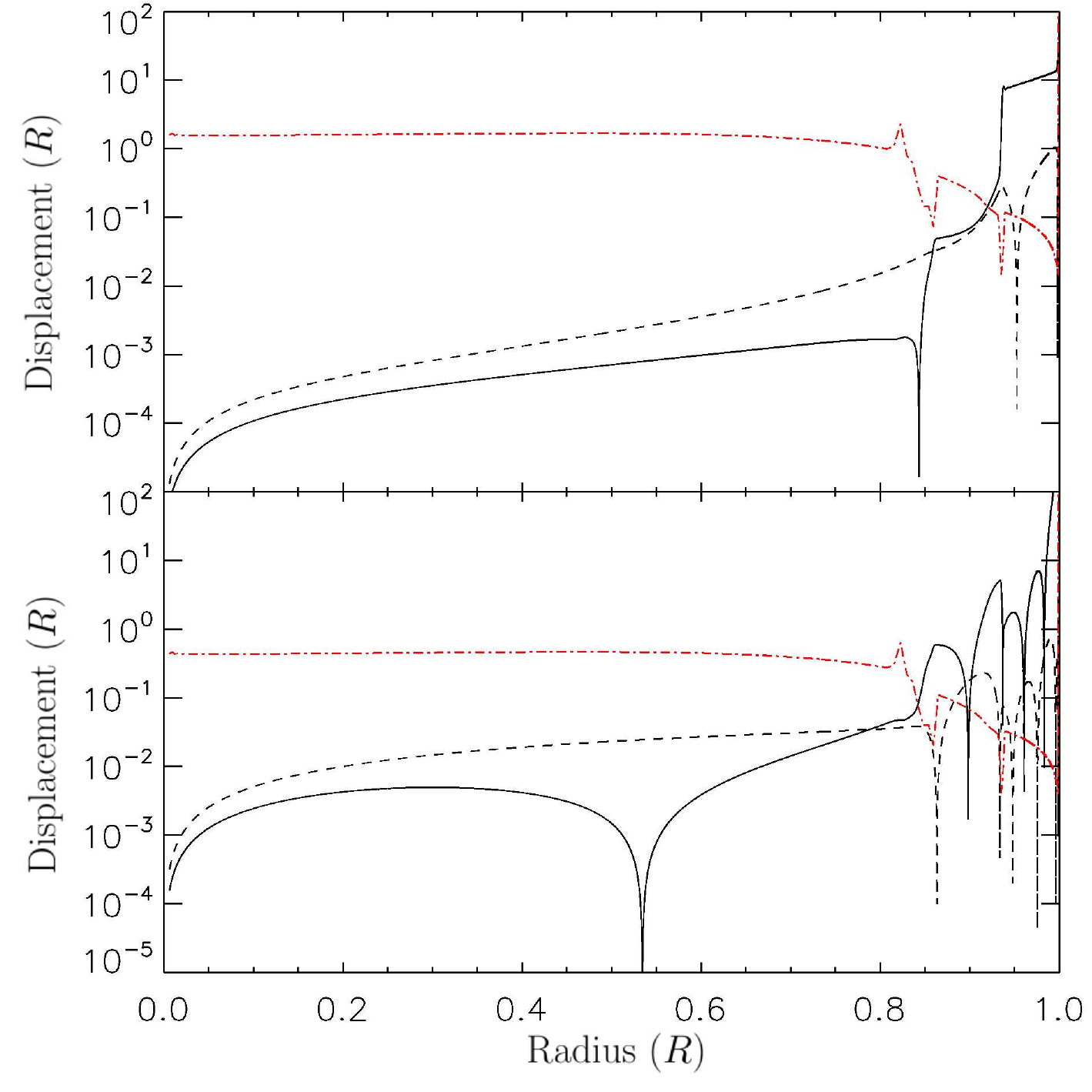}
\caption{\label{xi} The horizontal (solid line) and radial (dashed
  line) displacements of the $n=1$ (top panel) and $n=5$ (bottom
  panel) modes as a function of radius.  The physical displacements
  are calculated using the analytical estimates for the post-resonance
  amplitudes given in equation \ref{eq:bmax} using $M=M'$, giving $|a_1|=0.0566$ and $|a_5|=0.0695$.
 Also shown is inverse of the WKB wave number $1/k_r$ (dotted line).}
\end{centering}
\end{figure}

However, before these implications can be taken seriously, one should
be aware of the limitations of the present study. One issue is the
assumption that the white dwarf is non-rotating (and not
synchronized), already commented on in section 1. More importantly, we
have assumed that the white dwarf oscillations can be calculated in
the linear regime. While the mass-averaged dimensionless amplitude
$|a_\alpha|=|b_\alpha|$ of the excited g-mode is less than $0.1$ [see
Fig.~5 and Eq.~(\ref{eq:bmax})], the physical fluid displacement in the
stellar envelope is much larger since g-modes of white dwarfs are
mainly concentrated in the outer, non-degenerate layers.  Figure
\ref{xi} gives some examples: it shows the horizontal and radial
displacements of three modes at their post-resonance amplitudes. These
are obtained from $\bxi = a_\alpha \bxi_\alpha$, with $\bxi_\alpha$
the normalized eigenfunction (see section \ref{modes}) and $a_\alpha$
computed from equation (\ref{eq:bmax}) with $M'=M$. In general, the
linear approximation is valid only if $|\bxi| \ll |k_r|^{-1}$, where
$k_r$ is the WKB wave number, given by 
\be
\label{kr}
k_r^2 \simeq \frac{l(l+1)(N^2-\omega^2)}{\omega^2 r^2}.
\ee
Clearly, the three modes depicted in Fig.~8 strongly violate the linear
approximation beyond the radius $r \approx 0.85 R$, near the jump in $N^2$
associated with the carbon-helium boundary.

Therefore, the results presented in this paper should be treated with
caution as nonlinear effects will likely limit mode growth.  Rather
than increasing to the large displacements shown in Figure \ref{xi}, the
white dwarf oscillations will undergo non-linear processes such as
mode coupling that will transfer energy to high-order modes.
These high-order modes have much shorter wavelengths and thus damp on very
short time scales. As nonlinearity is most important in the outer layers
of the white dwarf, we expect that the excited 
oscillation will dissipate its energy preferentially
in this outer layer and will not reflect back into the stellar interior.
We plan to address these issues in our next paper (Fuller \& Lai 2010,
in preparation).

\section*{Acknowledgments}

We thank Gilles Fontaine (University of Montreal) for providing the white
dwarf models used in this paper and for valuable advice on these
models.  DL thanks Lars Bildsten, Gordon Ogilvie and Yanqin Wu for
useful discussions, and acknowledges the hospitality of the Kavli
Institute for Theoretical Physics at UCSB (funded by the NSF through
Grant PHY05-51164) where part of the work was carried out.
This work has been supported in part by NASA Grant NNX07AG81G and NSF
grants AST 0707628.

\def\apj{{Astrophys. J.}}
\def\apjs{{Astrophys. J. Supp.}}
\def\mnras{{Mon. Not. R. Astr. Soc.}}
\def\prl{{Phys. Rev. Lett.}}
\def\prd{{Phys. Rev. D}}
\def\apjl{{Astrophys. J. Let.}}
\def\pasp{{Publ. Astr. Soc. Pacific}}
\def\aapr{{Astr. Astr. Rev.}}


\end{document}